\documentclass[14pt, 
superscriptaddress,
aps,prx,
reprint,
twocolumn,
amsmath,amssymb,showpacs
]{revtex4-2}
\usepackage{titlesec}
\usepackage{newtxtext,newtxmath}
\usepackage{physics}
\usepackage{graphicx}
\usepackage{dsfont}
\usepackage{xcolor}
\usepackage{mathbbol}
\usepackage{amsmath}
\definecolor{urlc}{RGB}{58,105,157}
\usepackage[
colorlinks=true,
urlcolor=urlc,
linkcolor=urlc,
citecolor=urlc
]{hyperref}

\usepackage{algorithm}
\usepackage{algpseudocode}
\usepackage{algorithmicx}

\begin{document}   

\title{Speedup of high-order unconstrained binary optimization using
quantum $\mathbb{Z}_2$ lattice gauge theory}  
\author{Bi-Ying Wang}
\affiliation{Wilczek Quantum Center, Shanghai Institute for Advanced Studies, Shanghai 201315, China}
\affiliation{Yangtze River Delta Industrial Innovation Center of Quantum Science and Technology, Suzhou 215100, China}
 
\author{Xiaopeng Cui}
\email{xpclove@126.com}
\affiliation{Department of Physics \& State Key Laboratory of Surface Physics, Fudan University, Shanghai, 200433, China}

\author{Qingguo Zeng}

\affiliation{Shenzhen Institute for Quantum Science and Engineering, Southern University of Science and Technology, Shenzhen, 518055, China}

\author{Yemin Zhan}

\affiliation{Department of Physics \& State Key Laboratory of Surface Physics, Fudan University, Shanghai, 200433, China}

\author{Man-Hong Yung}
\email{yung@sustech.edu.cn}
\affiliation{Shenzhen Institute for Quantum Science and Engineering, Southern University of Science and Technology, Shenzhen, 518055, China}
\affiliation{International Quantum Academy, Shenzhen, 518048, China}
\affiliation{Guangdong Provincial Key Laboratory of Quantum Science and Engineering, Southern University of Science and Technology, Shenzhen, 518055, China}
\affiliation{Shenzhen Key Laboratory of Quantum Science and Engineering, Southern University of Science and Technology, Shenzhen, 518055, China}

\author{Yu Shi}
\email{yu\_shi@ustc.edu.cn}
\affiliation{Wilczek Quantum Center, Shanghai Institute for Advanced Studies, Shanghai 201315, China}
\affiliation{Department of Physics \& State Key Laboratory of Surface Physics, Fudan University, Shanghai, 200433, China}

\date{\today}

\begin{abstract}
An important and difficult problem in  optimization is the  high-order unconstrained binary optimization, which can represent many optimization problems more efficient than quadratic unconstrained binary optimization, but how to quickly solve it has remained difficult. Here we present an approach  by mapping  the high-order unconstrained binary optimization to quantum $\mathbb{Z}_2$ lattice gauge theory defined on the dual  graph, and propose the gauged local quantum annealing, which is the local quantum annealing protected by the gauge symmetry. We present the  quantum algorithm and its corresponding quantum-inspired classical algorithm for this problem, and achieve algorithmic speedup by using gauge symmetry.   
By running the quantum-inspired classical algorithm, we demonstrate that the gauged local quantum annealing reduces the computational time by one order of magnitude from that of  the local quantum annealing.
\end{abstract}

\maketitle

\section{Introduction}

High-order unconstrained binary optimization (HUBO)  and quadratic unconstrained binary optimization (QUBO) and  are  two high-performance fundamental binary models for the  important and wide-range problems of combinatorial optimization.  
With its simple quadratic-interaction formulation, QUBO has  been extensively studied over the past few decades. Numerous classical combinatorial optimization problems, such as Traveling Salesman Problem, Boolean Satisfiability Problem~\cite{frontier2014, kalComplex2022, IsingHardware2022}, Maximum Likelihood Detection Problem in the communication technology~\cite{kim2019leveraging}, error correction based on Low-Density Parity
Check, reconfigurable intelligent surfaces beamforming~\cite{ross_RIS_2022}, molecular unfolding~\cite{mato2022quantum} and protein folding problems~\cite{robert2021resource}, as well as the optimal path problem for routes~\cite{2022npjGlos}, have been successfully converted to QUBO problems.  
However, transformation of  these problems to QUBO problems necessitates lots of additional variables~\cite{xia2017electronic,boykov2004experimental,fujisaki2022practical}, as well as Rosenberg quadratization penalty terms~\cite{rosenberg1975reduction,QUBO_problem,qubo_prob,VERMA2022100594}, which  increase computational costs and make it challenging for the standard optimizations. 

In fact, these problems can be naturally expressed in terms of HUBO problems, where the cost functions are polynomials of orders higher than two. Therefore, instead of transforming them to QUBO problems,  solving them in HUBO formulation  can reduce both the number of binary variables and the difficulty in  model development, thus   save computational costs~\cite{mato2022quantum,2022npjGlos,blondel2016higher,ide2020maximum,bowles2022quadratic,norimoto2023quantum}.

Nevertheless,  QUBO  still has been used more widely than HUBO. The reason is that in QUBO, the quadratic-interaction formulation for binary variables maps to Ising model~\cite{barahona1982computational}, hence the tremendous amount of knowledge on Ising model has been useful in  the design of quantum algorithms~\cite{ebadi2022quantum,graham2022multi} and quantum inspired algorithms for QUBO~\cite{goto2019combinatorial, Ising_opti_EC, leleuScalingAdvantageChaotic2021, boothbyDWave2020}.  
Therefore, if Ising-like approaches are also established for HUBO, it is hopeful for HUBO to outperform QUBO.

It has been  noted that the Hamiltonian of quantum $\mathbb{Z}_2$ lattice gauge theory (QZ2LGT)~\cite{sachdev_Z2_2019a,wen_Z2_1991,wen_Z2_2002, hammaAdiabaticPreparationTopological2008}, with high-order interactions, can be studied by using the method of quantum simulation~\cite{cui_Z2_JHEP_2020} and maps to the Hamiltonian Cycle problem~\cite{cui_Z2_2022}. On the other hand,  gauge symmetry has already been used in quantum error detection, by measuring the conserved quantity with gauge operators  without disrupting the quantum evolution \cite{kitaev_QEC_2003a,zhang_QEC_2023,sivak_QEC_2023a}. 

In this paper, we  propose a method to map HUBO to QZ2LGT, which is regarded as 
a formulation of variable interaction in HUBO, and   
subsequently leverage  the gauge symmetry to improve HUBO solvers.   
A problem graph of  HUBO is  constructed \cite{cui_Z2_2022}, and QZ2LGT is defined on the dual. Afterwards, based on the gauge symmetry, a speedup scheme similar to quantum Zeno dynamics~\cite{zeno_2023,zeno_experimental_2014,facchi_zeno_2009} is introduced for computational speedup. The scheme is also adapted to  the corresponding quantum-inspired classical algorithms.

The gauge operators commute with the Hamiltonian. For the quantum adiabatic evolution, the time-dependent state is close to the instantaneous ground state, hence  the measurements   of the gauge operators enforce the reduction of the state to instantaneous ground state with high  probabilities, as the so-called quantum Zeno effect.  This feature is used in our quantum algorithm and the corresponding quantum-inspired classical algorithm.  We apply our method to upgrade the local quantum annealing (LQA) to the gauged local quantum annealing (gLQA). For comparison, we calculate the ground state energies of the QZ2LGT on a 2D lattice and on a  four-regular graph,  by using LQA, gLQA, and simulated annealing (SA) by levering  the capabilities of the advanced Python package OpenJij~\cite{Kirkpatrick1983,openjij}.  It is shown that gLQA outperforms LQA, which in turn outperforms SA.

\section{Methods}

\subsection{Mapping HUBO to $\mathbb{Z}_2$ gauge theory}
\label{sec:mapping}

We first discuss  how to map  HUBO to  QZ2LGT. Then the method to find out the gauge operators of the corresponding QZ2LGT is presented.

\begin{figure}[b]
    \centering
    \includegraphics[width=9cm]{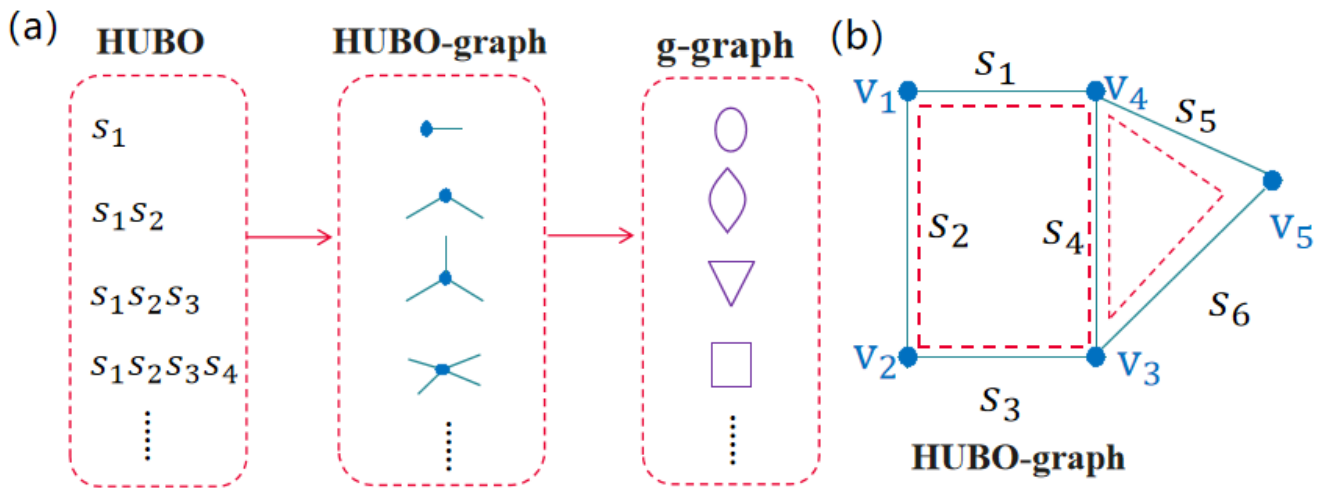}
    \caption{Procedure for mapping process. (a) Four different terms with interaction order varying from $1$ to $4$ in the objective function of HUBO map to four kinds of vertices in the \textit{HUBO-graph} and four kinds of plaquettes in the \textit{G-graph}. (b) An example of inefficient cycle in the \textit{HUBO-graph}. In this example, the cycle $s_1s_2s_3s_6s_5$ is inefficient, as it can be broken down into cycles $s_1s_2s_3s_4$ and $s_4s_5s_6$.}
    \label{fig:mapping}
\end{figure}

The objective of a HUBO is to minimize the classical Hamiltonian 
\begin{equation}
    H({\mathbf{s}}) = \sum_{i_1}J_{i_{1}}s_{i_1}+\sum_{{i_1<{i_2}}}J_{i_{1}i_2}s_{i_1}s_{i_2}+...+\sum_{{i_1<{i_2}<...<{i_{N}}}}J_{i_{1}i_2...i_{N}}\prod^{N}_{j=1}s_{j},
\end{equation}
for $N\geq 3$ with real-number coefficients $J$'s  and ${\mathbf{s}}\in\{-1,1\}^{N}$.

The procedure to map  HUBO to QZ2LGT involves two steps. First, we use a graph to describe the HUBO problem, which we refer to as the \textit{HUBO-graph}, in which each edge is occupied by a binary spin $s_i$ of the HUBO, while one vertex represents a term of the HUBO. Second, we map the \textit{HUBO-graph} to its dual, which we refer to as \textit{G-graph}. In this transformation, each edge in the \textit{HUBO-graph} is crossed by one link in the \textit{G-graph}, thus each vertex in the \textit{HUBO-graph} maps to a plaquette in the \textit{G-graph}, while two adjacent vertices in the \textit{HUBO-graph} map to two adjacent plaquettes in the \textit{G-graph}. 

To illustrate the procedure, we present the mapping from four different terms with interaction order varying from $1$ to $4$ in the objective function of HUBO  to four kinds of vertices of \textit{HUBO-graph} and four kinds of plaquettes of \textit{G-graph} in Fig.~\ref{fig:mapping}(a). As seen from the figure, each term in HUBO problem is represented as  a vertex in the \textit{HUBO-graph} and is subsequently mapped to a plaquette in the \textit{G-graph}. Finally, by placing spins at links of this \textit{G-graph}, we build   QZ2LGT   on the \textit{G-graph} and thus map the original HUBO to QZ2LGT, with the quantum  Hamiltonian 
\begin{equation}
    \hat{H} = \hat{Z} + g\hat{X},
    \label{eq:Ham_Z2}
\end{equation}
where $g$ is the coupling parameter and $\hat{X}=-\sum_l\hat{\sigma}_x^l$ denotes minus the sum of $\hat{\sigma}_x$ operations on all spins. $\hat{Z}=\sum_{p}{J_p}\prod_{l\in p}\hat{\sigma}^{l}_z$ represents the $\hat{\sigma}_z$ products operating on the spins of all plaquettes. The gauge operator which commutes with Hamiltonian $\hat{H}$ is defined as the product of $\hat{\sigma}_x$ on all links of each site $v$ 
\begin{equation}
    \hat{G}_v = \prod_{l\in{v}}\hat{\sigma}^l_x.
    \label{eq:z2_G}
\end{equation}

The ground state of the Hamiltonian of QZ2LGT is a superposition of product states of definite $\sigma_z$ of each spin. Each product state is an eigenstate of  $\hat{Z}$, and corresponds to an optimal solution to the classical HUBO problem.

\begin{figure}[b]
    \centering
    \includegraphics[width=7cm]{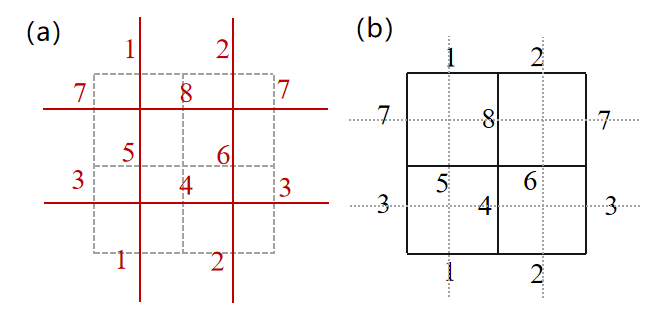}
    \caption{An example to illustrate mapping process. (a) \textit{HUBO-graph} and (b) \textit{G-graph} for HUBO with objective in Eq.~(\ref{eq:example}). Both graphs are obtained through the method introduced in Sec.\ref{sec:mapping}. The solid lines in each figure represent the edges of the graph, while the dashed lines indicate its dual graph. The \textit{G-graph} and the \textit{HUBO-graph} are the dual graphs for each other. The numbers denote the labels of binary spins in the HUBO problem.}
    \label{fig:dual_graph}
\end{figure}

The \textit{G-graph} is complicated, so it is not easy to obtain the gauge operators by directly counting the sites. Since each site on the \textit{G-graph} comes from an efficient cycle on  the \textit{HUBO-graph}, we apply the closed-loop search algorithm to identify these cycles directly in the HUBO-graph. An efficient cycle is the smallest cycle that cannot be further decomposed into smaller cycles. For instance,  in Fig.~\ref{fig:mapping}(b), the cycle $s_1s_2s_3s_6s_5$is inefficient, as it can be broken down into cycles $s_1s_2s_3s_4$ and $s_4s_5s_6$, which are efficient. Our algorithm is detailed in the Supplementary Methods. 

For  the mapping process, an example is given in the following. For a HUBO problem with an objective   
\begin{equation}
    H = J_1s_1s_3s_5s_4+J_2s_2s_4s_6s_3+J_3s_1s_8s_5s_7+J_4s_2s_7s_6s_8, 
    \label{eq:example}
\end{equation}
with eight variables $s_i$'s and four real-number coefficients $J$'s. As introduced above,  after mapping each item of the objective function into a vertex, the \textit{HUBO-graph} is obtained as in Fig.~\ref{fig:dual_graph}(a). Then, each edge of \textit{HUBO-graph} is crossed by a link, and the links around the same vertex should be connected to define the surrounded region as a plaquette in  the \textit{G-graph},  as shown in Fig.~\ref{fig:dual_graph}(b). As introduced in Eq.~(\ref{eq:z2_G}), the gauge operators for QZ2LGT defined in the \textit{G-graph} can be obtained by counting the sites. As seen in  Fig.~\ref{fig:dual_graph}(b),  with periodic boundary condition,  there are gauge operators 
$\hat{G}_1=\hat{\sigma}^4_x\hat{\sigma}^6_x\hat{\sigma}^8_x\hat{\sigma}^5_x$, $\hat{G}_2=\hat{\sigma}^1_x\hat{\sigma}^8_x\hat{\sigma}^2_x\hat{\sigma}^4_x$,$\hat{G}_3=\hat{\sigma}^3_x\hat{\sigma}^5_x\hat{\sigma}^7_x\hat{\sigma}^6_x$ and $\hat{G}_4=\hat{\sigma}^1_x\hat{\sigma}^2_x\hat{\sigma}^3_x\hat{\sigma}^7_x$.

In this paper, we will first propose quantum  speedup scheme for quantum adiabatic evolution based on the quantum Hamiltonian,  and then downgrade the  quantum Hamiltonian to classical Hamiltonian, and the speedup scheme is extended to  a classical one, as the quantum-inspired classical algorithm.

Before presenting our speedup scheme,  we review the previous scheme of LQA. 
In  quantum annealing, or adiabatic quantum computing,  one considers a system evolving under the time-dependent Hamiltonian 
\begin{equation}
    \hat{H}(t) = t\gamma\hat{H}_t-(1-t)\hat{H}_x, 
    \label{eq:ham_LQA}
\end{equation}
with $\gamma$ controlling the fraction of the energy of target Hamiltonian $\hat{H}_t$ in  the total Hamiltonian. The system is initially prepared in the state $|+\rangle^{\otimes{n}}$, which is the ground state of the Hamiltonian $-\hat{H}_x=-\sum_{i=1}^n\hat{\sigma}_x^i$, where $n$ is the number of spins in the system and $\hat{\sigma}_x^i$ is a Pauli operator on the $i$th spin.  The Hamiltonian varies from the initial Hamiltonian $-\hat{H}_x$ at $t=0$ to the target Hamiltonian $\hat{H}_t$ at time $t=1$.   If the variation speed of the Hamiltonian is slow enough to meet the adiabatic condition, the state of the system  stays at the instantaneous  ground state during the  evolution, reaching  the ground state of the target Hamiltonian   finally.

In LQA, which is inspired from quantum annealing, one only considers the states of the local form~\cite{bowles2022quadratic} 
\begin{equation}
    |\theta\rangle = |\theta_1\rangle\otimes|\theta_2\rangle\otimes...\otimes|\theta_n\rangle.
    \label{eq:lqa_theta}
\end{equation}
where $\theta_i$ denotes the angle between the state of the $i$th-spin with the z-axis, and is written in the form
\begin{equation}
    |\theta_i\rangle = \cos{\frac{\theta}{2}}|+\rangle+\sin{\frac{\theta}{2}}|-\rangle.
    \label{eq:lqa_theta_i}
\end{equation} 
The cost function of the LQA is defined as
\begin{equation}
     C(t,\boldsymbol{\theta}) = \langle\theta|\hat{H}(t)|\theta\rangle, 
    \label{eq:lqa_cost}
\end{equation} 
which can be written as a function of variable $\boldsymbol{\theta}$.
A variable $w_i\in{\mathbb{R}}$ is used to parameterize   $\theta_i$ as $\theta_i = \frac{\pi}{2}\tanh{w_i}$, in order to limit the range of $\theta_i$ to be $\theta_i\in[-\frac{\pi}{2},\frac{\pi}{2}]$. As a result, the cost function $C(t,{\mathbf{w}})$ is expressed as a function of variable $w_i(i =1,..,n)$ and time $t$.  

In the corresponding quantum-inspired classical scheme, with the time discretized as $t_j = j/N_{\rm iter} (j=1,..,N_{\rm iter})$, an iteration  of variables $w_i$, depending on the gradient of cost function is performed, 
$ \boldsymbol{\nu} \gets \mu{\boldsymbol{\nu}} - \eta\bigtriangledown_{\mathbf{w}}C({\mathbf{w}},t)$,  
${\mathbf{w}} \gets {\mathbf{w}} + \boldsymbol{\nu}$, 
where $\boldsymbol{\nu}$ is an additional vector representing the updating speed of ${\mathbf{w}}$ , $\mu\in[0,1]$ and  $\eta$ are parameters~\cite{bowles2022quadratic}. After the iteration, the final spin configuration can be obtained as $s_i = sign(w_i)$. 

\begin{figure*}[htbp]
    \centering
    \includegraphics[width=0.8\linewidth]{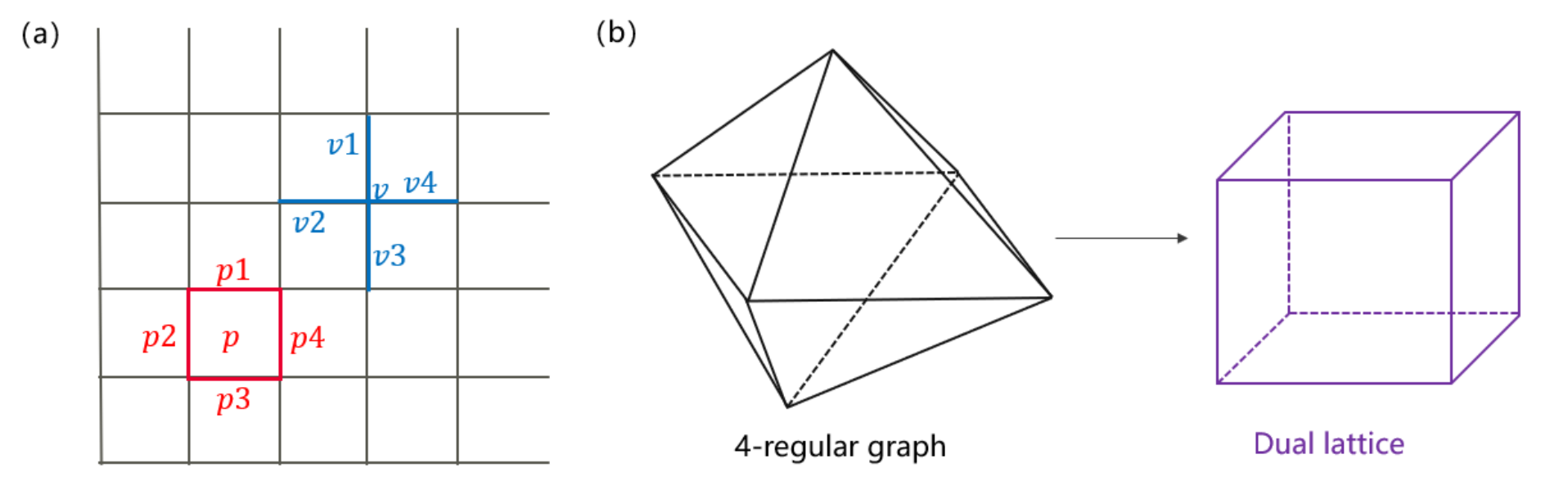}
    \caption{Examples to illustrate dual graphs. (a) The structure of a square lattice. The spins $p_{i} (i =1,2,3,4)$ represent the spins located on the edges of a plaquette, while the spins $v_{i} (i =1,2,3,4)$ correspond to the spins connected to a vertex. (b) The \textit{G-graph} obtained as the dual lattice of a four-regular graph.}
    \label{fig:lattice}
\end{figure*}

\subsection{Our speedup scheme and  the corresponding classical scheme}

During an adiabatic evolution,  as the evolution is not infinitesimally slow, the  state $|\psi(t)\rangle$  deviates from the instantaneous ground state $|\psi_g(t)\rangle$. A gauge operator $\hat{G}$ commutes the Hamiltonian $\hat{H}(t)$ at any time $t$, and it also commutes both $\hat{H}(0)=\hat{H}_x$ and $\hat{H}(1)=\hat{H}_t$. Therefore  $|\psi_g(t)\rangle$ is also an eigenstate of $\hat{G}$,
\begin{equation}
\hat{G}|\psi_g(t)\rangle = a|\psi_g(t)\rangle,
\end{equation}
with constant $a$. 

Our speedup scheme for quantum adiabatic evolution is the following. A  measurement  of the gauge  operator $\hat{G}$ is made on the state  $|\psi(t)\rangle$, and a  correction is subsequently made according to the measurement result $G$,  to force the system to its ground state. This process is repeated until the state $|\psi(t)\rangle$ becomes the ground state $|\psi_g(t)\rangle$.   

In  quantum simulation  with a finite step size, the adiabatic condition is not strictly met. Gauge symmetric  measurements can be employed to safeguard the adiabatic process. For a model with gauge symmetry, which should be preserved throughout the entire process, a  measurement  of the gauge operator does not interfere with the evolution. Thus, under the gauge symmetric measurements, the adiabaticity can be maintained within fewer steps, leading to speedup.

The above gauge-protected scheme for quantum adiabatic evolution can be extended  to a   classical speedup scheme with  similar protection. After transforming the problem to the classical binary optimization, the optimization Hamiltonian $H({\mathbf{s}})$ and symmetry operator  $G({\mathbf{s}})$ can be described in terms of  classical spin configuration $\mathbf{s} \equiv \{ s_i \}$. In our speedup  scheme, a symmetry-forced operation based on gradient is made,  on the spin configuration $\mathbf{s}$ in every step of the  algorithm, 
\begin{equation}
 s_i \gets s_i - B[G({\mathbf{s}})-a]\frac{\partial{G({\mathbf{s}})}}{\partial{s_i}},
\label{eq_speedup}
\end{equation}
where B is a parameter  controlling  the  evolution speed.  This is a classical feedback inspired by quantum measurement and quantum Zeno effect.

As an example of the above general scheme, we now present the formulation of gLQA by introducing gauge symmetry into LQA. During the quantum annealing process, the   state is always  the instantaneous ground state. Since each  gauge operator commutes with the system Hamiltonian, the state
in the adiabatic evolution is also an eigenstate of each  gauge operator. Thus, in LQA, 
\begin{equation}
    \hat{G}_{v_i}|\theta\rangle = |\theta\rangle,
\end{equation}
where $v_i$ represents a vertex of link $i$.  
As an example of the above classical feedback (\ref{eq_speedup}), after obtaining the localized classical formula of gauge operator $G_{v_i}$, 
an additional gradient-based iteration generated from the gauge operator is applied to force the state to respect the gauge symmetry, 
\begin{equation}
    {w_i} \gets {w_i} + B\sum_{v_i}(G_{v_i}-1)\frac{\partial{G_{v_i}}}{\partial{w_i}}, 
\end{equation}
where $v_i$ represents  the sites on link $i$, and $B$ is a constant. By replacing $G_{v_i}$ with $\prod_{l\in{v_i}}x_l$,  where $x_l \in [-1,1]$, the iteration for the gLQA becomes
\begin{eqnarray}
    \nu_i &\gets& \mu{\nu_i} - \eta\bigtriangledown_{w_i}C({\mathbf{w}},t) \nonumber\\
    {w_i} &\gets& {w_i} + {\nu_i} \nonumber\\
    {w_i} &\gets& w_i- B\sum_{v_i}(\prod_{l\in{v_i}}x_l-1)\frac{\partial{\prod_{l\in{v_i}}x_l}}{\partial{w_i}}.
    \label{eq:mlqa_iter}
\end{eqnarray}

\section{Results}

As introduced in Sec.~\ref{sec:mapping}, a HUBO task  can be mapped to the calculation of the ground state and its energy of  QZ2LGT on a \textit{G-graph}. In order to benchmark our algorithm, we apply LQA and gLQA to calculate the ground energies  of QZ2LGT on  two kinds \textit{G-graph}, namely, 2D square lattice and the dual graph of a random four-regular graph. In the first subsection, the structures of the two graphs are introduced. Then the calculation results of each graph are presented, which are performed on an Intel® CPU i7-8700 operating at a frequency of 3.2GHz. To benchmark the speed of algorithms, we introduce a standard merit -- time to solution (TTS),  which is defined as the computation time of finding an optimal value or solution with $99\%$ probability~\cite{2019SATTS,aramon2019physics}. TTS can be calculated through the real computation time $t_p$ as 
\begin{equation}
     {\rm TTS} = t_p\frac{\log(1-0.99)}{\log(1-p)},
     \label{eq:TTS}
\end{equation}
where success probability 
\begin{equation}
p \equiv \frac{n_{\rm sol}}{n_{\rm sam}}
\end{equation}
denotes the ratio between the number of times $n_{\rm sol}$ achieving the ground state energy and  the total number of sampling times $n_{\rm sam}$ with the corresponding computation time $t_p$. 

\subsection{2D lattice and four-regular graph}
\label{sec:graph}

On a square lattice of size $L\times{L}$,  as a dual graph, there are $2L^2$ links and $L^2$ plaquettes, and one spin locates at each link, so altogether there are $N=2L^2$ spins, $N_p=L^2$ plaquettes and $N_v=L^2$ vertices. The lattice is depicted in Fig.~\ref{fig:lattice}(a). The Hamiltonian of $\mathbb{Z}_2$ lattice gauge theory is as in Eq.~(\ref{eq:Ham_Z2}).  As seen from the figure, each magnetic term contains $\hat{\sigma}_z$ operating on the four spins ($p1,p2,p3,p4$) of one plaquette with $\hat{Z}=-\sum_{p}\hat{\sigma}^{p1}_z\hat{\sigma}^{p2}_z\hat{\sigma}^{p3}_z\hat{\sigma}^{p4}_z$, while the gauge operator $\hat{G}_v$ contains four links of site $v$ and can be written as $\hat{G}_v=\hat{\sigma}^{v1}_x\hat{\sigma}^{v2}_x\hat{\sigma}^{v3}_x\hat{\sigma}^{v4}_x$. 

\begin{figure}[b]
    \centering
    \includegraphics[width=9cm]{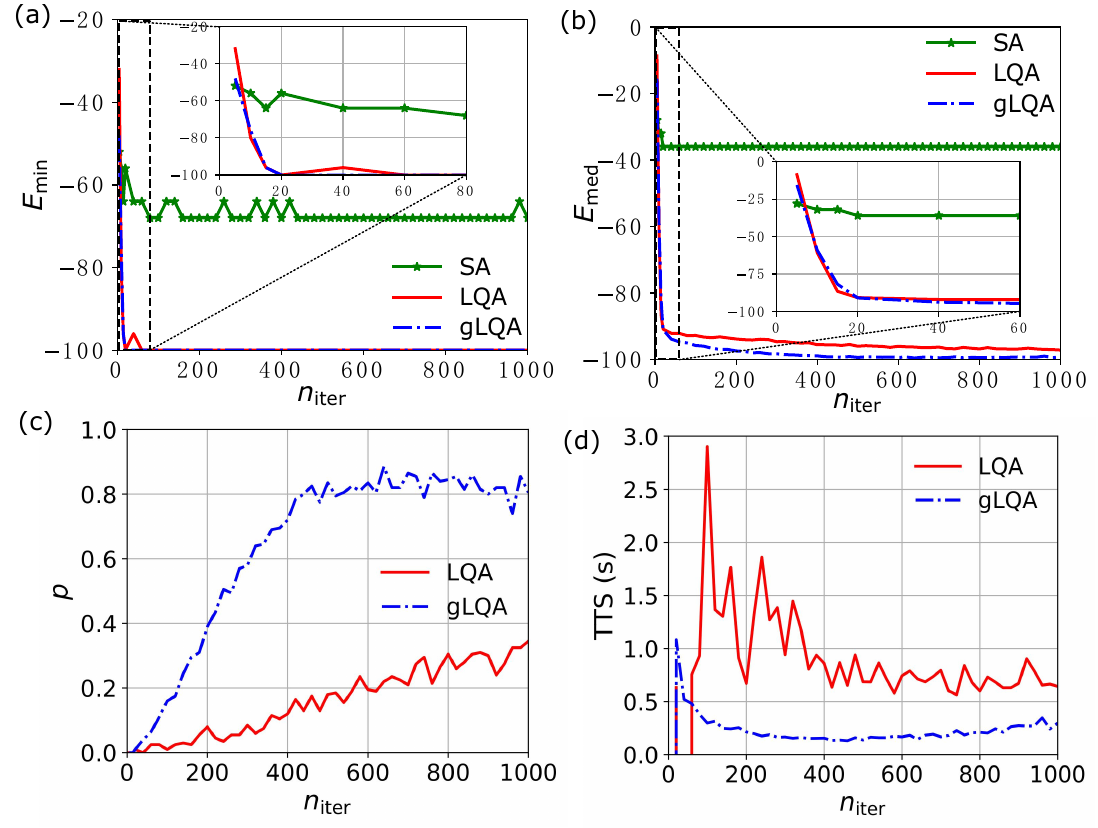}
    \caption{
For  a square lattice with $L = 10$,  
     (a) minimum energy $E_{\rm min}$, (b) medium energy $E_{\rm med}$, (c) success probability $p$, (d) time to solution (TTS), as functions of iteration steps $n_{\rm iter}$.    The results obtained from three algorithms   (LQA, gLQA and SA) are all based on sampling number  $n_{\rm sam}=200$. The insets in (a) and (b) provide an enlarged view of the data for small iteration steps in the corresponding graph.}
    \label{fig:L=10}
\end{figure}

We generate a random \textit{G-graph} by finding the dual lattice of a random four-regular graph as presented in  Fig.~\ref{fig:lattice}(b). As seen from the figure, every edge in the four-regular graph maps to one link in the \textit{G-graph} which crosses the edge, and every vertex or site   in the  four-regular graph maps to a plaquette in the \textit{G-graph},   an efficient cycle in the four-regular graph maps to a site in the \textit{G-graph}. As explained in  Sec.\ref{sec:mapping}, by counting all the vertices, the operator $\hat{Z}$ and gauge operator $G_v$ of the \textit{G-graph} can be obtained. In the four-regular graph, each vertex has four connected edges, which consequently lead  to four links $\{{p1,p2,p3,p4}\}$ in every plaquette $p$ of the obtained \textit{G-graph}. Thus, the $\hat{Z}$ operator can be  written as $\hat{Z}=-\sum_{p}\hat{\sigma}^{p1}_z\hat{\sigma}^{p2}_z\hat{\sigma}^{p3}_z\hat{\sigma}^{p4}_z$. Due to the uncertain length of the cycle  in a randomly generated four-regular graph, the number of links in one site $s$ of \textit{G-graph} is uncertain. For convenience,  in the following, we only collect the gauge operators for which the number of spins included is smaller that a threshold $k_m$ for the \textit{G-graphs} generated by four-regular graphs.

\subsection{Numerical results for 2D lattices}

In this subsection, we present our gLQA results of  optimization for a square lattice, as the dual graph. We first  discuss the results for the lattice size $L=10$, then  present the results for a range of lattice sizes from 10 to 40. For comparison, we also include the results of LQA optimization. 
For  $L=10$,   optimization  from SA is also presented.

A lattice contains $N_p=L^2$ plaquttes, so the ground state energy is $E_s=-N_p$. For $L=10$, the minimum energy $E_{\rm min}$ and median energy $E_{\rm med}$, which is the median value  of the energy over all the samples, as functions of the number of iteration steps $n_{\rm iter}$, for sampling number  $n_{\rm sam}=200$, are depicted in   Fig.~\ref{fig:L=10}(a) and Fig.~\ref{fig:L=10}(b), respectively.  
As can be seen from Fig.~\ref{fig:L=10}(a) and Fig.~\ref{fig:L=10}(b), though  $E_{\rm min}$ and $E_{\rm med}$ obtained from SA (green curve in the figures)  exhibit rapid decreases  and subsequent saturations   with the increase of iteration steps $n_{\rm iter}$, they  fail to converge to the ground state values within 200 times of sampling. 
As seen from Fig.~\ref{fig:L=10}(a), as   $n_{\rm iter}$ increases,   $E_{\rm min}$ obtained from LQA and  that from gLQA rapidly decrease  and saturate  towards the ground state energy $E_s=-100$ within $60$ steps of iteration. Moreover,   $E_{\rm min}$ obtained from gLQA reaches the ground state  energy quicker than from  LQA. As seen from  Fig.~\ref{fig:L=10}(b),  similar behavior can be observed for   $E_{\rm med}$ calculated from  LQA and that  from gLQA, which  decreases  rapidly before gradually slows  down until   saturation. Besides,   $E_{\rm med}$ obtained from gLQA is smaller  than that from LQA. The probability $p$ and TTS as functions of iteration step $n_{\rm iter}$ are displayed in Fig.~\ref{fig:L=10}(c) and \ref{fig:L=10}(d) respectively.  As seen from Fig.~\ref{fig:L=10}(c), the probability $p$ calculated from gLQA and that from LQA gradually increase with $n_{\rm iter}$, leading to the decrease of TTS in the Fig.~\ref{fig:L=10}(d). For $n_{\rm iter}\geq 500$, the probability $p$ calculated in gLQA saturates to a high value $\sim 0.81$, inducing a slow increase of TTS with $n_{\rm iter}$. The probability $p$ calculated in LQA saturates to $p\sim0.35$ for $n_{\rm iter}\geq 1000$. It is clearly seen in  Fig.~\ref{fig:L=10}(d) that gLQA achieves shorter TTS than LQA. Notably, the shortest TTS of approximately 0.15~s in gLQA for this lattice occurs at $n_{\rm iter} = 500$, which is less than a quarter of the shortest TTS of approximately $0.60~s$ for LQA, demonstrating a fourfold increase of the speed by our gLQA.

\begin{figure}[b]
    \centering
    \includegraphics[width=9cm]{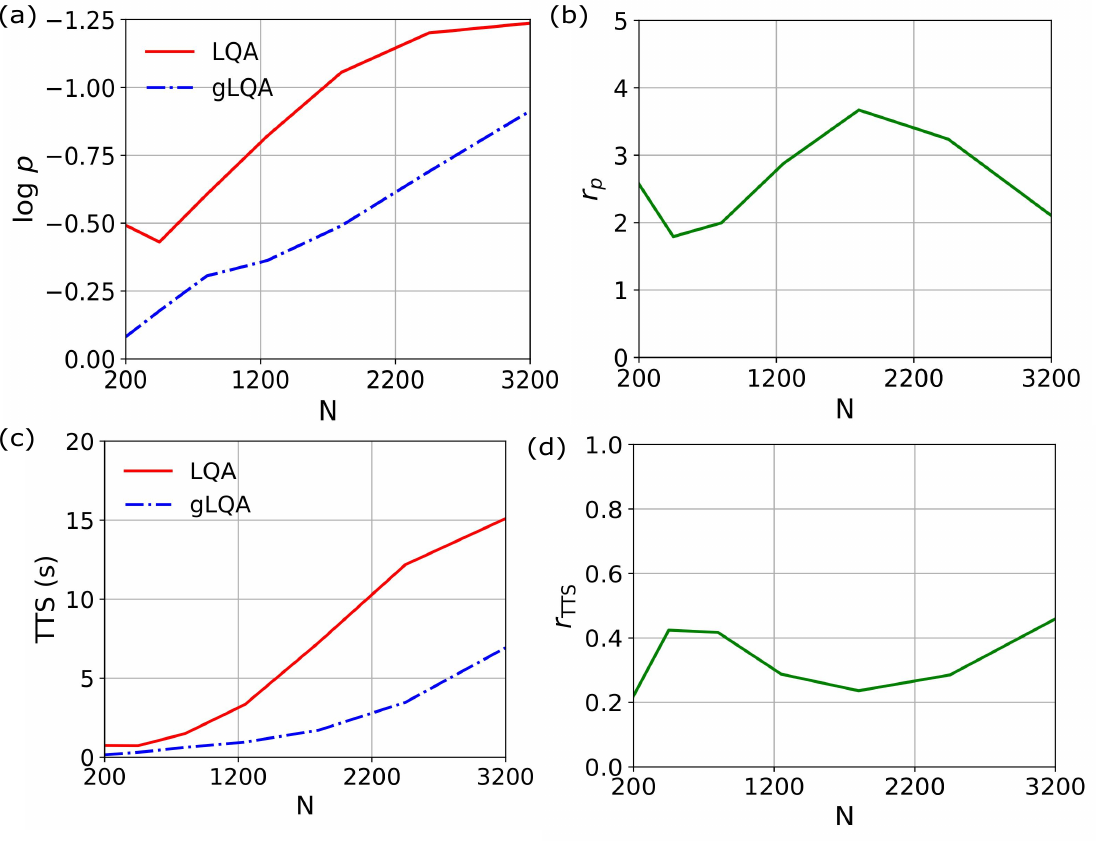}
    \caption{(a) Success probability $p$, (b) success possibility ratio $r_p$ of algorithm gLQA to algorithm LQA, (c) time to solution TTS, (d) time-to-solution ratio $r_{\rm TTS}$ of algorithm gLQA to algorithm LQA,  as functions of   number of spins $N$. The red bold curves and blue dotted-dashed curves in (a) and (c) denote the results  from the LQA and gLQA algorithms, respectively. The green bold curves in (b) and (d) represent  the ratio of results between gLQA and LQA.}
    \label{fig:L=10_40}
\end{figure}

The   results with $n_{\rm sam}=1000$ for lattices with various  values of  number of spins $N$ ($L$ ranging from $10$ to $40$) are illustrated in Fig.~\ref{fig:L=10_40}.  
Since the results obtained from SA cannot reach the ground state for a small lattice with $L=10$, we only compare  LQA and gLQA.  
With the increase of the number of spins  $N$,  for the iteration step fixed at $1000$,  the success probability $p$ in  Fig.~\ref{fig:L=10_40}(a) decreases, which consequently leads to the increased TTS in the Fig.~\ref{fig:L=10_40}(c). To make a clear comparison, the $N$ dependence of ratio $r_p$ ($r_{\rm TTS}$) between the success probability (TTS) calculated in  our gLQA and that in LQA is presented in Fig.~\ref{fig:L=10_40}(b) (Fig.~\ref{fig:L=10_40}(d)). The figures clearly indicate that our gLQA method achieves at least a twofold increase in success probability $p$ and  a reduction of TTS by $60\%$.
\begin{figure}[b]
    \centering
    \includegraphics[width=9cm]{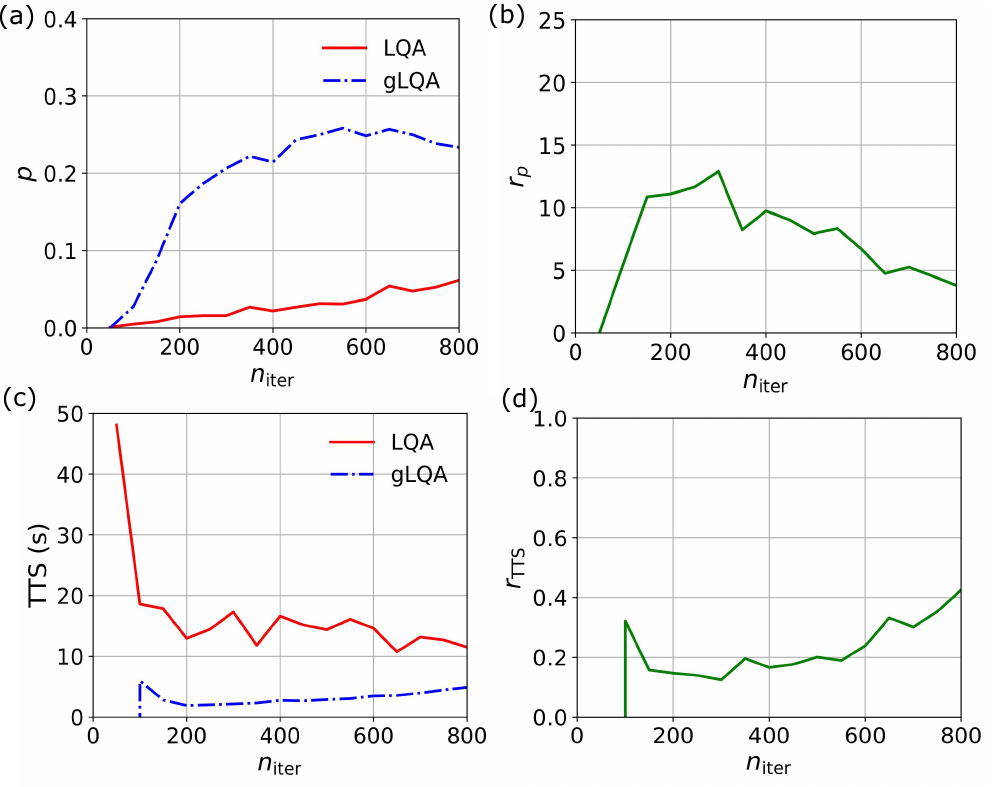}
    \caption{For the \textit{G-graph} as the dual of  a random four-regular graph with   number of spins  $N = 400$,  (a) success probability $p$, (b) success possibility ratio $r_p$ of algorithm gLQA to algorithm LQA, (c) time to solution TTS, (d) time-to-solution ratio $r_{\rm TTS}$ of algorithm gLQA to algorithm LQA,  as functions of iteration steps $n_{\rm iter}$ . The red bold curves and blue dotted-dashed curves in (a) and (c) denote  the results obtained from the LQA and gLQA algorithms, respectively. The green bold curves in (b) and (d) represent  the ratio of results between gLQA and LQA.}
    \label{fig:4r200}
\end{figure}

\subsection{Results for \textit{G-graphs} dual to four-regular graphs}

In this subsection, we present our results for \textit{G-graphs} generated as the dual from four-regular graphs. We first discuss the results for a \textit{G-graph} with $N=400$. Then, the outcomes for  $N$ ranging from 400 to 1600 are presented. 

 The  success probability and TTS are calculated as functions of the number of iteration steps $n_{\rm iter}$, for   $n_{\rm sam}=2000$ samplings of  a random \textit{G-graph} with $400$ links, as  shown in   Fig.~\ref{fig:4r200}(a) and Fig.~\ref{fig:4r200}(c). As seen from Fig.~\ref{fig:4r200}(a), with the increase of $n_{\rm iter}$, the success probability $p$ in   LQA and that in gLQA  both start  to increase gradually from $0$ for $n_{\rm iter} > 100$. Then  the success probability $p$ in  gLQA saturates around $0.25$ for $n_{\rm iter}\geq500$, while $p$ in  LQA only reaches $\sim0.05$ for $n_{\rm iter}=800$. As $p=0$ for $n_{\rm iter}<100$,  TTS in this regime is meaningless.  As seen from  
 Fig.~\ref{fig:4r200}(c), with the increase of $n_{\rm iter}$, TTS’ in both LQA and gLQA first  decrease, because of the increase of  $p$.
 It is notable that the shortest TTS in LQA is around 11s, which is more than fivefold of the one in our gLQA, which is  $\sim2~$s. The   ratios $r_p$ and $r_{\rm TTS}$ as   functions of $n_{\rm iter}$ are  shown in the Fig.~\ref{fig:4r200}(b) and Fig.~\ref{fig:4r200}(d),    respectively.  Fig.~\ref{fig:4r200}(b) displays  at least a fourfold increase of the success probability by  our gLQA, which  reaches thirteenfold at $n_{\rm iter}=300$.   As shown in Fig.~\ref{fig:4r200}(d), TTS in our gLQA has a $\sim60\%$ reduction compared with the one from LQA, which   reaches $\sim85\%$ at $n_{\rm iter}=300$. 
 
\begin{figure}[b]
    \centering
    \includegraphics[width=9cm]{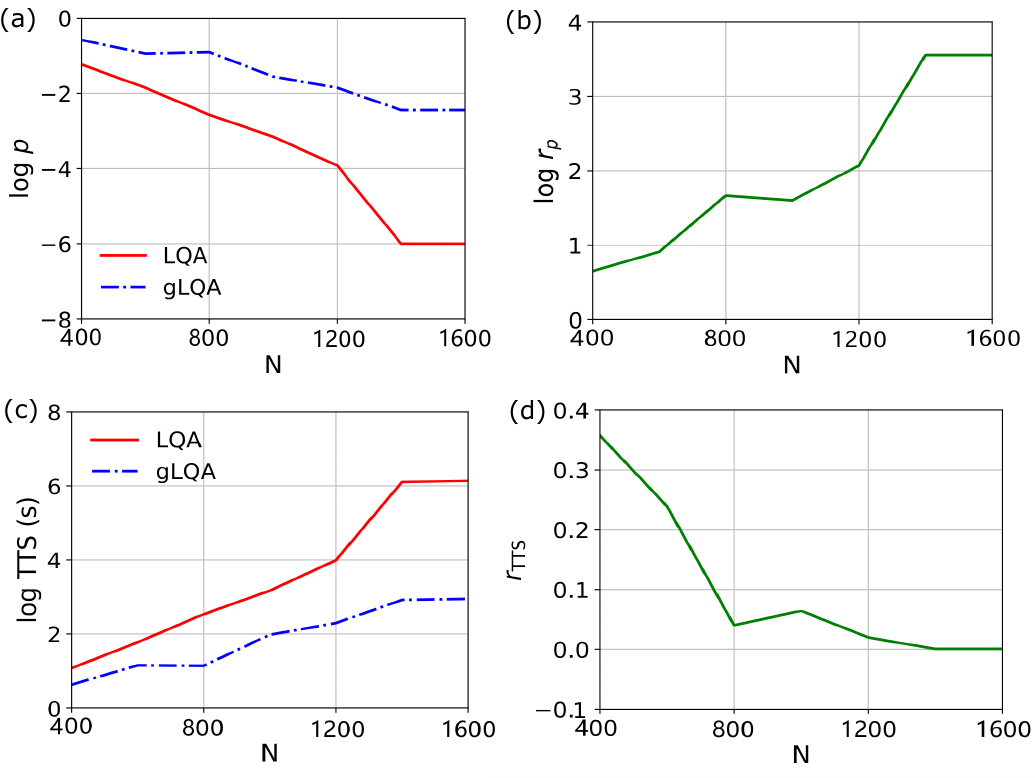}
    \caption{For  the \textit{G-graph}s as the duals of random four-regular graphs,  (a) success probability $p$, (b) success possibility ratio $r_p$ of algorithms gLQA to LQA, (c)  time to solution (TTS), (d) time-to-solustion ratio $r_{\rm TTS}$ of algorithms gLQA to LQA,  as functions of the  number of spins  $N$. The red bold curves and blue dotted-dashed curves in (a) and (c) denote the results obtained from the LQA and gLQA algorithms, respectively. The green bold curves in (b) and (d) represent the ratio of results between gLQA and LQA.}
    \label{fig:4r200_800}
\end{figure}

 The  results with $n_{\rm sam}=100000$ for \textit{G-graph} generated from four-regular graphs with   the number of spins  $N$ ranging from $400$ to $1600$   are presented in  Fig.~\ref{fig:4r200_800}.  With the increase of $N$, for the iteration step kept  constant at $500$,  the success probability $p$ decreases, as shown in   Fig.~\ref{fig:4r200_800}(a), and consequently leads to the increase of TTS, as shown in  Fig.~\ref{fig:4r200_800}(c). Besides, a sudden increase of TTS in  LQA appears  for $N>1200$  due to the sudden decrease of near-zero success probability $p$. To make a clear comparison,  $N$ dependences of ratio $r_p$ and $r_{\rm TTS}$ are presented in Fig.~\ref{fig:4r200_800}(b) and Fig.~\ref{fig:4r200_800}(d). The figures clearly indicate that our gLQA method achieves an obvious increase in   $p$, while  displays a reduction of TTS larger than $75\%$. For large \textit{G-graph} with $N>1200$, the increase of success probability for our gLQA can be larger than $3000$ and the reduction of TTS can reach three orders of magnitude.

\section{Conclusions}

In this paper,  we first presented a quantum algorithm for HUBO by mapping it to a graph problem, which is subsequently transformed to QZ2LGT defined  on the dual  graph, referred to as the \textit{G-graph}.  The   HUBO problem is thus transformed to the problem of finding the ground state and its energy  in QZ2LGT. The gauge operators commute with the Hamiltonian, hence their measurements  enforce the state to be in ground state during its evolution,  leading to the  speedup of the adiabatic algorithm.  

Then we  presented the corresponding  quantum-inspired  classical algorithm  facilitated by the gauge symmetry, and have introduced a gauge-forced iteration step to further speed up the computation.   

We have demonstrated the advantage of our method by  introducing  gLQA in the quantum algorithm and the corresponding quantum-inspired algorithm. 

The benchmarking analyses have been made on the quantum-inspired algorithms using LQA and gLQA,  on two types of \textit{G-graphs} with spin numbers ranging from 200 to 3200. Our benchmarking analyses of TTS based on LQA and gLQA demonstrated that gLQA significantly outperforms LQA.  
For comparison, we have also applied SA to solve these problems,  which fails even for a small 2D lattice of size $L = 10$.

We have successfully identified  $\mathbb{Z_2}$ gauge theory as suitable for HUBO and leveraged gauge symmetry to expedite the solution process.   

There are several potential avenues for future developments.  First, it would be beneficial to test our speedup scheme on a wider range of instances, and develop an automated technique for tuning the parameters in our scheme, such as constant B and the time step, which are determined through preliminary searches now.

Moreover, the proposed gauge-symmetry-protected scheme has the potential to accelerate computation for all algorithms based on quantum adiabatic theory.
It is also interesting to explore whether other  features of QZ2LGT are useful in this scheme.    

{\bf Data Availability} The data generated and/or analyzed in this work are available from the corresponding author upon reasonable request.

{\bf Code Availability} The codes generated in this work are available from the corresponding author upon reasonable request.

{\bf Acknowledgements}
This work was supported by National Natural  Science Foundation of China (Grant No. 12075059).

{\bf Author Contributions}
B.Y.W., X.C. and Y.S. conceived the work and formulated the theoretical framework. B.Y.W. performed the numerical calculations and analyzed the data in useful discussion with X.C., Y.S. and M.H.Y..  B.Y.W., X.C., Q.Z.,  Y.Z. and Y.S. prepared the manuscript. All authors contributed to analyzing the data, discussing the results and commented on the writing.

{\bf Competing Interests}
The authors declare no competing interests.

\section*{References}
\bibliography{ref}

\end{document}